\documentclass[iop]{emulateapj}

\newcommand{\D}{\ensuremath{\mathcal{D}}}

\newcommand{\Msun}{\ensuremath{M_{\odot}}}

\newcommand{\be}{\begin{equation}}
\newcommand{\ee}{\end{equation}}

\shorttitle{Diffusive Nuclear Burning of Helium on NSs}
\shortauthors{Chang, Bildsten, \& Arras}

\begin{document}

\title{Diffusive Nuclear Burning of Helium on Neutron Stars} 

\author{Philip Chang\altaffilmark{1} 
Lars Bildsten\altaffilmark{2,3} and Phil Arras\altaffilmark{4}}

\altaffiltext{1} {Canadian Institute for Theoretical Astrophysics,
  University of Toronto, 60 St George St, Toronto, ON M5S 3H8,
  Canada; pchang@cita.utoronto.ca}

\altaffiltext{2}{Kavli Institute for Theoretical Physics, Kohn Hall,
  University of California, Santa Barbara, CA 93106;
  bildsten@kitp.ucsb.edu}

\altaffiltext{3}{Department of Physics, Broida Hall, University of
  California, Santa Barbara, CA 93106}

\altaffiltext{4}{Department of Astronomy, University of
  Virginia, Charlottesville, VA 22904; pla7y@virginia.edu}

%\date{Accepted 1988 December 15. Received 1988 December 14; in original form 1988 October 11}

\begin{abstract}
  Diffusive nuclear burning of H by an underlying material capable of
  capturing protons can readily consume H from the surface of neutron
  stars (NSs) during their early cooling history.  In the absence of
  subsequent accretion, it will be depleted from the photosphere.  We
  now extend diffusive nuclear burning to He, motivated by the recent
  observation by Ho \& Heinke of a carbon atmosphere on the NS in the
  Cassiopeia A supernova remnant.  We calculate the equilibrium
  structure of He on an underlying $\alpha$ capturing material,
  accounting for thermal, mass defect, and Coulomb corrections on the
  stratification of material with the same zeroth order $\mu_e = A/Z$. We show that Coulomb corrections dominate over thermal and mass defect corrections in the highly degenerate part of the envelope.  We also show
  that the bulk of the He sits deep in the envelope rather than near
  the surface.  Thus, even if the photospheric He abundance is low,
  the total He column could be substantially larger than the photospheric column, which may have
  implications for rapid surface evolution ($\approx 1$ yr timescales)
  of neutron stars.  When nuclear reactions are taken into account, we
  find that for base temperatures $\gtrsim 1.6
  \times 10^8$ K, He is readily captured onto C.  As these high
  temperatures are present during the early stages of NS evolution, we
  expect that the primordial He is completely depleted from the NS
  surface like the case for primordial H.  We also find that magnetic
  fields $\lesssim 10^{12}$ G do not affect our conclusions.  Armed
  with the results of this work and our prior efforts, we expect that
  primordial H and He are depleted, and so any observed H or He on the
  surfaces of these NS must be due to subsequent accretion (with or
  without spallation).  If this subsequent accretion can be prevented,
  the underlying mid-Z material would be exposed.
\end{abstract}

\keywords{diffusion -- 
nuclear reactions, nucleosynthesis, abundances -- pulsars: general -- stars:
abundances, interiors -- stars: neutron}

\section{Introduction}\label{sec:intro}

The surface composition of neutron stars (NSs) remains an outstanding
problem.  Observations of young NSs have failed to find indisputable
signatures of any particular element on their surface.  Some NSs
completely lack any spectral features such as RX J1856.5-3754 (Burwitz
et al. 2001, 2003; Pons et al. 2002; Ho et al 2007).  However, when
spectral features are seen, the degeneracies in our (currently)
limited understanding of the chemistry and physics of highly
magnetized material makes secure identification a challenge.  Theory
also fails to constrain the initial conditions, i.e., supernova
fallback, spallation, subsequent accretion, that would make a
convincing case for a particular composition.

Even after the initial composition is set, subsequent evolution can
occur.  For instance, the small mass ($\approx 10^{-17}\Msun$) and rapid
diffusion time ($\approx 1$ s) at the NS photosphere suggest that it
should consist of H due to gravitational settling.  However, we showed
in Chang \& Bildsten (2003,2004; hereafter paper I and II) and Chang,
Arras, \& Bildsten (2004; hereafter paper III) that H is easily
destroyed by diffusive nuclear burning (DNB).  The central idea behind
DNB, which was first proposed by Chiu \& Salpeter (1968; see also
Rosen 1970), is that H can diffuse to great depth where the
temperature and density are sufficiently large to consume H by proton
captures onto heavier elements.  In paper I and II, we showed that
this process is so effective that we expect NS surfaces to be depleted
of any primordial H.  The observation of H on the surfaces of NS would
then point to late-time or continuous accretion.  This conclusion is
insensitive to the strength of the magnetic field and the size of an
inert He buffer that sits between the H and the underlying proton
capturing elements (paper III).

With all the H consumed, one would expect surfaces of He.  However,
recent observations by Ho \& Heinke (2009) suggest that the NS at the
center of the Cassiopeia A supernova remnant has a carbon surface with
an effective temperature $T_{\rm e} = 1.8\times 10^6\,{\rm K}$ and
radius $R = 12-14$ km.  This paper shows that He is also vulnerable to
diffusive nuclear burning on the surface of Cas A.  All primordial
H/He is consumed during its early cooling history, exposing the
underlying material. If subsequent accretion does not cover this
underlying material, we would expect a population of NSs with mid-Z
surfaces.

We first review the physics of DNB in \S\ref{sec:physics}, outlining
the major results of papers I-III.  We then discuss the physics of
material in diffusive equilibrium on NSs in \S\ref{sec:efield}.  We
show how to calculate the electric field in degenerate material
accounting for thermal, mass defect, and Coulomb corrections.  Using this electric
field, we then calculate the structure of an envelope composed of He
and C. We show that the diffusive tail of He penetrates sufficiently
deep to be readily captured by C, and find that He is easily consumed
in young, hot NSs for a range of $\alpha$-capturing elements.  In
\S\ref{sec:implications}, we conclude with a discussion of the
implications of our work for young NS surface compositions.

\section{Basic Physics for Hydrogen Depletion}\label{sec:physics}

The diffusion rates on NSs imply that H will rapidly float to
the photosphere where the density is $\rho\approx 1\,{\rm g\,cm}^{-3}$
and the temperature of $T\approx 10^6$ K.  Hydrogen at this density
and temperature will not burn in a Hubble time.  However, $10^3-10^4$
cm below the photosphere, the higher densities ($\rho > 10^6\,{\rm
  g\,cm}^{-3}$) and temperatures ($T\sim 1-10\times 10^8\,{\rm K}$)
allow H to burn.  The H that diffuses down to this depth will be
readily captured, driving a H current from the surface to the burning
layer. Over time, this results in the total depletion of H.

In paper I, we studied the case where the rate of H burning is
sufficiently slow that H always remain in diffusive equilibrium, and
the rate limiting step is the rate at which H is consumed by nuclear
reactions in the burning layer. This burning layer is defined by the
competing effects of an exponential decline in H number density and a
rising temperature.  For cooler NSs (with effective temperatures $T_e < 10^6\,{\rm
  K}$), DNB takes place in this nuclear-limited regime.  For hotter
NSs ($T_e \ga 1.5 \times 10^{6}\,{\rm K}$), DNB is not limited by
nuclear reactions in the burning layer, but rather by the rate at
which H can diffuse down to the burning layer, i.e., the diffusion
limited regime (paper II).  In this case, the burning occurs in a thin
layer at which the nuclear burning time becomes comparable to the
downward diffusion time, breaking the
assumption of diffusive equilibrium (see papers I and II).  Below this burning
layer, the hydrogen concentration profile is rapidly cut off.

In papers I and II, we demonstated that DNB can easily consume all the
H on NS surfaces early in their cooling history.  We also studied the
effect of magnetic fields on the rate of DNB and found that for fields
$\lesssim 10^{12}$ G, this basic evolution is unaffected.  In paper
III, we extended our study of DNB to magnetar surfaces.  There, we
showed that the high temperatures associated with magnetars, i.e.,
soft gamma repeaters and anomalous X-ray pulsars, allow H to be
captured onto elements as heavy as Fe.  In addition, we showed that
the effect of an inert He buffer between the H and C does little to
damp the effectiveness of DNB.  Our calculation of the structure of an
(assumed) inert He on C in paper III was an early attempt at
calculating the structure of a degenerate He/C atmosphere (including
Coulomb physics).  We refine our methodology as we describe below.

\section{Electric Field in Degenerate Plasma}\label{sec:efield}

We first discuss the compositional profile of degenerate material in
diffusive equilibrium, set by the competition between gravity and the
electric field.  To simplify the discussion initially, we consider an
ideal gas of electrons and ions. In section 3.2, we include non-ideal
effects. The equations of
hydrostatic balance of a plasma of ions and electrons in a plane
parallel atmosphere are
\begin{eqnarray}\label{eq:hb ions} 
\frac {\partial P_i} {\partial r} &=& -n_i \left( A_i m_p g - Z_i e E\right), \\ 
\frac {\partial P_e} {\partial r} &=& -n_e e E,\label{eq:hb electrons}
\end{eqnarray}
where $n_i$ ($n_e$) is the number density of ions (electrons), $P_i$
($P_e$) is the ion (electron) pressure, $A_i$ and $Z_i$ are the mass
and charge of the ion, $g=GM_{\rm NS}/R_{\rm NS}^2$ is the gravitational acceleration, $M_{\rm NS}$ and $R_{\rm NS}$ are the mass and radius of the NS, and $E$ is the electric field.  
Note that we ignore the electron mass, and we set the nuclear mass to $Am_p$, where $m_p$ is set to be the atomic mass unit.  This small change, which differs from our definitions in papers I, II, and III, allows us to include mass defect corrections.  However, by defining the atomic mass unit as $m_p$ rather than the more standard $u$, we maintain consistent nomenclature with papers I, II, and III. 
The electric field in equation (\ref{eq:hb
  ions}) and (\ref{eq:hb electrons}) results from the condition of
charge neutrality $n_e = Z_i n_i$.  For an ideal, nondegenerate plasma
($P_e = Z P_i$), this electric field is $eE = A_im_pg/(Z_i + 1)$.  For
a strongly degenerate plasma ($P_e \gg P_i$), the electric field becomes
\begin{equation}\label{eq:zeroth order E-field}
  eE = \frac {A_i}{Z_i} m_p g.
\end{equation}
If we now consider the forces on a trace ion species j immersed in a
background ion species i, equation (\ref{eq:hb ions}) becomes
\begin{equation}
\frac {\partial P_j} {\partial r} = -n_j
m_p g\left( A_j - Z_j \frac {A_i}{Z_i}\right).
\end{equation}
For $A_j/Z_j = A_i/Z_i$, the differentiating force between trace and
background ion species is zero for the zeroth order electric field of
equation (\ref{eq:zeroth order E-field}).  Note this is not the case
of H on C in paper I, II, and III, which have very different $\mu_e =
A/Z$ and so the zeroth order electric field imposes a differentiating
force.  Here ions with the same $A_i/Z_i$ experience no
differentiating force (ignoring mass defect corrections) in a strongly degenerate
plasma.  Hence, higher order corrections to the electric field due to
thermal effects and Coulomb interactions determine the differentiating
forces between ion species and the equilibrium compositional
structure.

\subsection{Thermal, Mass Defect, and Coulomb Corrections}\label{sec:thermal}

We now give a qualitative discussion of the thermal and Coulomb
corrections to the electric field in degenerate plasmas.  We first
consider corrections to the zeroth order electric field due to purely thermal
effects because this physics is simpler and illustrative.  Using
equations (\ref{eq:hb ions}) and (\ref{eq:hb electrons}) and the
condition of charge neutrality, the electric field is (paper I),
\begin{eqnarray}\label{eq:E-field cb03}
  eE &=& \left[ \sum_i n_i Z_i \left(A_im_p g + k_B
  \left({\partial T}/{\partial z}\right)\right) - \right.\nonumber \\ 
  &&\left.\left({\partial P_e}/{\partial T}\right)\left({\partial
  P_e}/{\partial n_e}\right)^{-1} \left({\partial T}/{\partial z}\right)\right] \nonumber \\ &&\over { \sum n_iZ_i^2 +
  n_e \left({n_e k_B T}/{P_e}\right) \left( {\partial\ln P_e}/{\partial\ln
  n_e}\right)^{-1}},
\end{eqnarray}
where $z = r - r_{\rm ref}$ is the height measured from a reference
point, $r_{\rm ref}$.  Note that we have ignored the mass of the electron in the derivation of equation (\ref{eq:E-field cb03} (see paper I for a detailed derivation). As expected, we reproduce the zeroth order
electric field (eq.[\ref{eq:zeroth order E-field}]) from equation
(\ref{eq:E-field cb03}) for a degenerate plasma.

Now consider a two component plasma where $n_1 \gg n_2$ (1 is the
background and 2 is the trace) with the same $\mu_e$, i.e., $A_1/Z_1
= A_2/Z_2$.  For a degenerate plasma ($P_e \gg n_e k_B T$), we expand
equation (\ref{eq:E-field cb03}) in the ratio of the thermal pressure
to the degenerate pressure, $n_{e,1} k_B T/P_e$, to find
\begin{equation}\label{eq:E-field thermal}
eE = \frac {A_1} {Z_1} m_p g\left( 1 - \frac {n_1 k_B T} {\gamma P_e}\right),
\end{equation} 
where $\gamma = {\partial\ln P_e}/{\partial\ln n_e}$, demonstrating that the thermal
correction yields a differentiating force between ions with the same
$A_i/Z_i$.  We should note that Hameury, Heyvaerts and Bonazzola (1983) derived 
a similar results.

Plugging equation (\ref{eq:E-field thermal}) into hydrostatic balance
for each ion, we find
\begin{eqnarray}
\frac {\partial \ln P_1} {\partial r} = -\frac {\rho g}{\gamma P_e}, \nonumber \\ 
\frac {\partial \ln P_2} {\partial r} = -\frac {Z_2} {Z_1} \frac {\rho g}{\gamma P_e},
\end{eqnarray}
where we presume that the ions obey the ideal gas equation of state,
$P_i = n_i k_B T$.  Subtracting these equations from each other and
using hydrostatic balance $\partial P/\partial r = -\rho g$, where $P \approx P_e$, we
find a simple power law between the concentration, $n_2/n_1$, and
pressure, $P$,
\begin{equation}\label{eq:sameA/Zlaw}
\frac {\partial\ln n_2/n_1} {\partial \ln P} = \gamma^{-1}\left(\frac {Z_2}{Z_1} -
1\right).
\end{equation}
Equation (\ref{eq:sameA/Zlaw}) demonstrates that thermal corrections
to the electric field yield a differentiating force which separates
lighter ions from heavier ions with the same $A_i/Z_i$ in a degenerate
plasma.  

Our discussion of purely thermal effects above, while illustrative ignores mass defect corrections and Coulomb corrections.  Mass defect corrections arise from that fact that $A_j/Z_j $ is only $\approx A_i/Z_i$ for different $i$ and $j$ due the the slight variation in atomic mass.  While this is usually ignored, as we have done in paper I, II, and III, the anonymous referee has reminded us that mass defect corrections are important in the degenerate plasmas at this level of approximation that we are interested in (see Blaes et al. 1992).  Fortunately this is an easy fix as our derivation of equation (\ref{eq:E-field cb03}) is independent of any assumption regarding A and Z of the ions so including mass defect corrections merely requires including the appropriate atomic masses of the ions.\footnote{We have used the atomic database compiled by National Institute of Standards and Technology available at http://www.nist.gov/physlab/data/comp.cfm} However, we will show below that Coulomb corrections ultimately dominate over both thermal and mass defect corrections.

The alert reader may worry that these mass defect corrections may change the results of paper I, II, and III.  However, he or she may rest assured that these corrections are orders of magnitude smaller than the order unity corrections of H on C which we discussed in paper I, II, and III.  The only possible exception may be in regard to our discussion of the He buffer in paper III.  However as we will show below, Coulomb effects continue to dominate, which argues that our results on the He buffer in paper III remains valid.   

\begin{figure}
%  \epsscale{0.9}
  \plotone{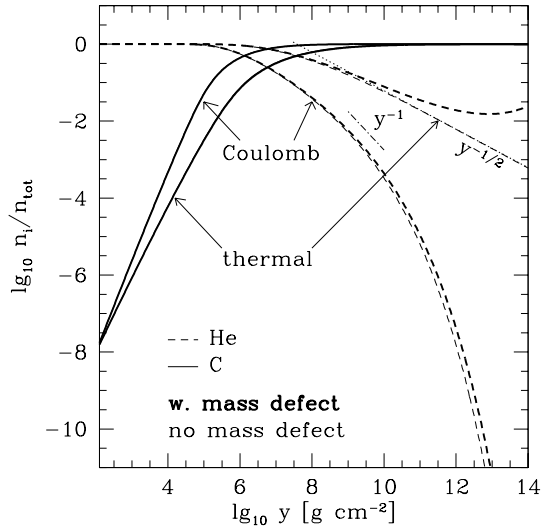}
  \caption{ Fractional abundance of He (dashed lines) and C (solid
    lines) as a function of $y$ in an equilibrium atmosphere with $T_e
    = 10^6$ K and $g = 2.43\times 10^{14}\,{\rm cm\,s}^{-2}$. The thin (thick)
    lines labeled thermal only include thermal (and mass defect) corrections and the thin (thick) lines
    labeled Coulomb contain both thermal and Coulomb corrections (thick lines also include mass defect corrections), although
    Coulomb effects dominate over both thermal effects and mass defect corrections. The dotted line is a $y^{-1/2}$ power
    law (eq. [\ref{eq:sameA/Zlaw}]) for a relativistic electron
    equation of state ($\gamma = 4/3$) and this analytic result tracks
    the exact numerical calculation for purely thermal corrections. The dot-dashed line denotes a
    $y^{-1}$ power law. The respective He columns, $y_{\rm He} = \int\rho_{\rm He} dz$, where $\rho_{\rm He}$ is the local He density, are $y_{\rm He} \approx 4\times 10^{10}\,{\rm g\,cm}^{-2}$ and $\approx 7\times 10^{11}\,{\rm g\,cm}^{-2}$ for purely thermal corrections and for thermal and mass defect corrections, respectively and $y_{\rm He} \approx 1.2\times 10^{7}\,{\rm g\,cm}^{-2}$ and $\approx 1.5\times 10^{7}\,{\rm g\,cm}^{-2}$ for thermal + Coulomb corrections and including mass defect corrections, respectively.
}
  \label{fig:HeCstructure}
\end{figure}

Figure \ref{fig:HeCstructure} plots the He (dashed lines) and C (solid
lines) as a function of column depth, $y = \int \rho dz$, for an
equilibrium atmosphere with temperature, $T = 10^6$ K.  Note that $y = P/g$ in a
plane parallel, constant gravity atmosphere.  In thin lines, we show the He and C
abundance accounting for only thermal corrections to the electric
field (thin lines marked by thermal) and accounting for both thermal and
Coulomb corrections (thin lines marked by Coulomb; see \S \ref{sec:coulomb} for a discussion on how these Coulomb corrections are calculated). In thick lines, we also include the mass defect corrections to the masses of He and C.  

The analytic power law (eq. [\ref{eq:sameA/Zlaw}] for $\gamma = 4/3$ (dotted line marked by
$y^{-1/2}$) matches that given by a full numerical calculation
for He for purely thermal corrections  (thin dashed line marked thermal).  
Including mass defect corrections results in a significant change (thick dashed line marked thermal).  The mass defect correction drives the abundance of He away from both the analytic power law and numerical calculation for purely thermal corrections (dotted line marked by $y^{-1/2}$ and thin dashed line marked thermal, respectively).
In fact, the abundance of He increases rather than decreasing at large column. To understand this, we note that the mass defect correction for He on C is
\begin{equation}
  \frac{\Delta F_{\rm defect}}{m_p g} = (A_{\rm He} - 2 Z_{\rm He}) \approx 0.003,
\end{equation}
where $A_{\rm He}$ is the corrected mass number of He.  Thermal corrections, using equation (\ref{eq:E-field thermal}), are 
\begin{equation}
  \frac{\Delta F_{\rm Th}}{m_p g} = Z_{\rm He} \frac{A_{\rm C}}{Z_{\rm C}} \frac{n_{\rm C} k_B T}{\gamma P_e} \approx 0.002,
\end{equation}
when the plasma is highly degenerate.  The mass defect correction is fixed relative to $m_p g$ for increasing column because it depends only on the mass numbers of He and C. On the other hand, the thermal correction scale like $n_{\rm C}k_BT/P_e$, which decreases for increasing column.  Morever, the nucleons in He are heavier than the nucleons in C, so the larger mass defect correction opposes the thermal correction, reversing the He abundance trend at large depth.   

Coulomb effects, however, overwhelm both the thermal and the mass defect corrections, which we also show in Figure \ref{fig:HeCstructure} in the lines marked Coulomb.  A quick order of magnitude estimate of the size of this Coulomb correction is (see the discussion in \S \ref{sec:coulomb} especially around eqn. [\ref{eq:simple E-field}] for details) 
\begin{equation}
  \frac{\Delta F_{\rm Col}}{m_p g} = \frac {3Z_{\rm He}}{10(\gamma-1)} \frac {Z_{\rm C}^{2/3} e^2/a_e}{E_F} \approx 0.006
\end{equation}
for $y=10^{10}\,{\rm g\,cm}^{-3}$ and $g = 2.43\times 10^{14}\,{\rm cm\,s}^{-2}$ and assuming that $n_{\rm C} \gg n_{\rm He}$.  Clearly Coulomb corrections dominate over both the thermal and mass defect corrections as the effects of these forces on the number density of ions is exponential.  However, we have included all these effects in subsequent calculations in the remainder of this paper. 

Figure \ref{fig:HeCstructure} also shows that the fraction of the He
trace declines against the heavier C background, but the rate at which
this fraction declines strongly depends on whether the corrections to
the electric field include thermal effects, mass defect effects, or
Coulomb effects.  The dot-dashed line is for an abundance that varies
inversely with column, i.e., $n_i/n_{\rm tot} \propto y^{-1}$.  For He
abundances that decline shallower that this, the total He column
$y_{\rm He}$, rises substantially with increasing depth.  For He
abundances that decline steeper than $y^{-1}$, $y_{\rm He}$ is capped
and does not increase substantially with increasing $y$.  If we only
include thermal corrections to the electric field, the He abundance
follows the power law (eq. [\ref{eq:sameA/Zlaw}]), which is shallower
than $y^{-1}$. Hence, $y_{\rm He}$ continually rises with increasing
$y$.  Adding the mass defect effect to the thermal correction only make $y_{\rm He}$ increase faster.
However, if we include Coulomb physics, we find that at sufficient depth
($y \gtrsim y_{\rm cut} \approx 4 \times 10^{9}\,{\rm g\,cm}^{-2}$),
the abundance is steeper than $y^{-1}$ and hence the He column does
not increase substantially for increased $y$.  Below $y\lesssim y_{\rm
  cut}$, however, the He abundance is shallower than this and so the
He column continuously increases.  Hence, we find that the {\it bulk
  of the He sits at a depth of $y_{\rm cut}$}, which we show
explicitly below.  This will have important implications on the rate
of He DNB as we show in \S\ref{sec:He DNB}.

\subsection{Coulomb Effects via a Chemical Equilibrium Approach}\label{sec:coulomb}

Now we turn our attention to the effect of Coulomb physics on the
electric field.  While it would be tempting to follow the derivation
leading to equation (\ref{eq:E-field cb03}), but substitute
Coulomb correction terms for the ion pressure, this gives erroneous
results in the general case.  Instead we have calculated these
corrections starting from the condition of chemical equilibrium. De
Blasio (2000) performed a similar calculation to study the interface
between two dissimilar ions near the crust of a neutron star.  Our
calculation below is similar in spirit, but we make contact between
the various methods of calculating the corrections to the electric
field, i.e., starting from the hydrostatic balance for each ion
(eq.[\ref{eq:hb ions}]) vs. chemical equilibrium. The condition of
chemical equilibrium is equivalent to the equation of hydrostatic
balance for each species, but has the advantage that the effects of
Coulomb physics is more naturally incorporated.

We begin by considering the free energy of an ideal plasma, which is 
\begin{equation}\label{eq:ideal free energy}
F = F_{\rm id,deg}^{(e)}(N_e,V,T) + F_{\rm id,nondeg}^{(i)}(N_i,V,T),
\end{equation}
where $F_{\rm id,deg}$ and $F_{\rm id,nondeg}^{(i)}$ are the ideal
free energies of the degenerate electrons and nondegenerate ions, and
$N_e$ and $N_i$ are the total number of electrons and ions in a
volume, $V$.  For a nondegenerate ideal gas, $F =
-Nk_BT(\ln(V/N\lambda_i^3) + 1)$, where $\lambda_i =
\sqrt{2\pi\hbar^2/m_ik_BT}$ is the thermal wavelength.  From the free
energy, we find the chemical potential via
\begin{equation}
\mu^{(i)} \equiv \left(\frac{ \partial F}{\partial
N_i}\right)_{N_{j\ne i}, T, V}.
\end{equation}
Thus, the chemical potential for a nondegenerate ideal gas is
$\mu^{(i)} = \mu_{id}^{(i)} = k_BT\ln\left(n\lambda_i^3\right)$, where $n = N/V$ is
the number density.  In the presence of external fields, the chemical
potential becomes $\mu = \mu_{id} + \Phi$, where $\Phi$ is the potential
associated with the external field (Landau \& Lifshitz 1980).  For
gravitational and electric fields, the external potential is
\begin{equation}
\Phi = z(A_i m_p g - Z_i eE),
\end{equation}
where, again $z = r - r_{\rm ref}$.  Imposing chemical equilibrium
$d\mu/dz = 0$, we find for an isothermal, nondegenerate, plasma,
\begin{eqnarray}\label{eq:HB} 
kT \frac {\partial \ln n_i} {\partial r} = -A_i m_p g + Z_i e E, \\
kT \frac {\partial \ln n_e} {\partial r} = e E,
\end{eqnarray}
which recovers the equation of hydrostatic balance for each species
(in the isothermal limit) and demonstates the relationship between
chemical equilibrium and hydrostatic balance.

To consider Coulomb physics, we include an additional
term to the free energy:
\begin{equation}
F  =  F_{\rm id,deg}^{(e)}(N_e,V,T) + F_{\rm id,nondeg}^{(i)}(N_i,V,T) +
F_{ex}(N_e,N_i,V,T), 
\end{equation}
where $F_{ex}(N_e,N_i,V,T)$ is the excess free energy due to Coulomb
interactions.  Chabrier \& Potekhin (1998) give an excellent discussion of the Coulomb
physics of electron-ion plasmas and we adopt their formalism
and notation.\footnote{Chabrier \& Potekhin (1998)'s results are suitable at this level of approximation, but see Potekhin \& Chabrier 2000 and Potekhin,
Chabrier, \& Rogers 2009 for more accurate results.} Following Chabrier \& Potekhin (1998), we rewrite the
excess free energy $F_{ex} = N_e f_{ee} k_BT + N_i\left(f_{ii} +
f_{ie}\right)k_BT$ in terms of dimensionless functions,
$f_{ee}(\Gamma_e)$, $f_{ii}(\Gamma_e)$, and $f_{ie}(\Gamma_e)$, which
denote the electron-electron, ion-ion and screening interactions
respectively.  We define $\Gamma_e = e^2/a_ek_B T$ as the electron coupling
parameter, and $a_e = (4\pi n_e/3)^{-1/3}$ as the mean electron
spacing.  The ion coupling parameter $\Gamma_i =
Z^2e^2/a_ik_BT$ is related to $\Gamma_e$ via $\Gamma_i =
Z_i^{5/3}\Gamma_e$, where $a_i = (4\pi n_i/3)^{-1/3}$.

Typically the ion-ion term ($f_{ii}$) dominates and for the purposes
of simplifying the discussion, while capturing the relevant physics,
we consider just this term.  The chemical potentials then become
\begin{eqnarray}
\mu^{(e)} & \equiv & \left. \frac{ \partial F}{\partial N_e} \right|_{N_i,V,T}
= \mu^{(e)}_{id} + \frac{k_B T}{3} \sum_i \frac{N_i}{N_e} u_i, \\
\mu^{(i)} & \equiv & \left. \frac{ \partial F}{\partial N_i} 
\right|_{N_e,N_{j\neq i},V,T} = \mu^{(i)}_{id} + k_B T f_i,
\end{eqnarray}
where we have defined $f_i=f_{ii}(\Gamma_i)$ and $u_i = \partial
f_{ii}(\Gamma_i)/\partial \ln \Gamma_i$.  Chabrier \& Potekhin (1998)
(also see Potekhin et al 2009) have provided fitted forms for $u_i$
(their eq. [27]) which we use below.

Equilibrium requires $\partial\mu/\partial r=ZeE-Am_pg$ for each species.
Ignoring temperature gradients, we find
\begin{eqnarray}
\frac{\partial\ln n_i}{\partial r}  =  - \frac{u_i}{3} \frac{\partial \ln n_e}{\partial r}
+ \frac{Z_i eE - A_i m_p g}{k_B T},\label{eq:chemical equilibrium ions}\\
\left.\frac{\partial \mu^{(e)}_{id}}{\partial \ln n_e}\right|_T \frac{\partial \ln n_e}{\partial r}
+ \frac{1}{3} k_B T \sum_i \frac{n_i u_i}{n_e} 
\nonumber \\
\times\left[ \frac{\partial\ln n_i}{\partial r}+ \frac{\partial\ln n_e}{\partial r}(-1+h_i/3) \right]
 =  -eE, \label{eq:chemical equilibrium electrons}
\end{eqnarray}
where $h_i \equiv \partial\ln u_i/\partial\ln \Gamma$. For a single ion species,
$n_e = Z_i n_i$ and therefore $d\ln n_i = d\ln n_e$.  So equations
(\ref{eq:chemical equilibrium ions}) and (\ref{eq:chemical equilibrium
  electrons}) simplify to give
\begin{eqnarray}
k_B T\left(1 + \frac {u_i} 3\right)\frac{\partial\ln n_i}{\partial r} = Z_i eE -
A_i m_p g,\label{eq:ion eq}\\
\left(\left.\frac{\partial \mu^{(e)}_{id}}{\partial \ln n_e}\right|_T  + \frac{1}{9} k_B T \frac{u_i h_i}{Z_i}\right) \frac{\partial\ln n_e}{\partial r} = -eE. \label{eq:electron eq}
\end{eqnarray}
Including Coulomb physics introduces corrections to both the ion
equation (\ref{eq:ion eq}) and the electron equation (\ref{eq:electron
  eq}).  The additional correction to the electron equation would have
been missed had we tried to derive the electron equation from the
hydrostatic equation (\ref{eq:hb electrons}).\footnote{The ion
  equation (\ref{eq:ion eq}) is also different from what we would have
  derived had we started from the ion hydrostatic equation (\ref{eq:hb
    ions}).}  The total pressure is $P
= - \left.\left(\partial F/\partial V\right)\right|_T$.  Therefore,
the non-ideal, Coulomb part is $P_{\rm Coul} = nkTu_i/3$, which agrees
with the sum of equations (\ref{eq:ion eq}) and (\ref{eq:electron
  eq}).

This is further complicated in the general case of multiple species
where the relation $d\ln n_i = d\ln n_e$ does not hold.  For instance,
consider a trace ion species against a background $Z_i n_i \ll n_e$,
equation (\ref{eq:chemical equilibrium ions}) becomes
\begin{equation}
\frac {\partial P^{\rm (id)}_i} {\partial r} = -n_i\left(A_i m_p g - Z_i e E + \frac 1
  3 k_B T u_i\frac {\partial \ln n_e} {\partial r}\right),
\end{equation}
where $P^{\rm (id)}_i = n_i k_B T$ is the ideal gas equation of state.
Here a simple substitution of a corrected ions pressure, $P_i = n_i
k_B T (1 + u_i/3)$, would have given erroneous results.  Distinct
terms must be included for both the ion and electron momentum
equations. The Coulomb interaction must not be regarded as a pressure
for each ion species, but rather as a force that is sensitive to the
electron gradient which is set by the background.

This discussion of these two simple cases (single ion species and
trace) highlights the problems with calculating the electric field
starting from the equations of hydrostatic balance for ions and
electrons (eq.[\ref{eq:hb ions}] and [\ref{eq:hb electrons}]) when
Coulomb physics plays a role.  Fortunately, our approach from the
viewpoint of chemical equilibrium avoids these problems.

We now solve the general case by first plugging the ion equation into
the Coulomb term of the electron equation. Collecting terms, the
result can be written in the intuitive form
\begin{equation}
\frac{\partial\ln n_e}{\partial r}  \equiv  \frac{Z_e^* eE - A_e^* m_p g}{E_e^*},
\end{equation}
where
\begin{eqnarray}
E_e^* & = & \left.\frac{\partial \mu^{(e)}_{id}}{\partial \ln n_e}\right|_T
+ \frac{1}{3} k_B T \sum_i \frac{n_i u_i}{n_e}
\left( \frac{h_i-u_i}{3} - 1 \right),
\\
Z_e^* & = & - 1 - \frac{1}{3} \sum_i \frac{n_i u_i Z_i}{n_e},
\\
A_e^* & = & - \frac{1}{3} \sum_i \frac{n_i u_i A_i}{n_e}.
\end{eqnarray}
This result can now be used to simplify the ion equation:
\begin{equation}
\frac{\partial\ln n_i}{\partial r}  \equiv  \frac{Z_i^* eE - A_i^* m_p g}{k_B T},
\end{equation}
where
\begin{eqnarray}
Z_i^* & = & Z_i - \frac{k_B T}{E_e^*} \frac{u_i Z_e^*}{3},
\\ 
A_i^* &  = & A_i - \frac{k_B T}{E_e^*} \frac{u_i A_e^*}{3}.
\end{eqnarray}
In this form, it is as if the charge and mass of each species (ions
and electrons) are shifted due to Coulomb interactions.  Using charge
neutrality, we solve for the electric field and find
\begin{eqnarray}\label{eq:full E-field}
\frac{eE}{m_p g} & = & \frac{ \sum_i ({n_i}/{n_e}) Z_i A_i^* - 
({k_B T}/{E_e^*}) A_e^*}{ \sum_i ({n_i}/{n_e}) Z_i Z_i^* 
- ({k_B T}/{E_e^*}) Z_e^* }.
\end{eqnarray}
One can easily show this equation has the correct limits for
noninteracting particles, for degenerate/nondegenerate and
nonrelativistic/relativistic electrons.

As an illustrative exercise, we use equation (\ref{eq:full E-field})
to find the electric field for a single ion species:
\begin{eqnarray}
\frac{eE}{m_p g} & = & \frac{A \left[ (\gamma-1)E_F + k_B T u h / 9 Z \right]}
{ Z \left[ (\gamma-1)E_F + k_B T u h / 9 Z \right] 
+ k_B T \left( 1 + u/3 \right)}
\nonumber \\ & \simeq & 
\frac{A}{Z} \left( 1 + \frac{3}{10(\gamma-1)}\frac{Z^{2/3}e^2/a_e}{E_F} \right)
\label{eq:simple E-field}
\end{eqnarray}
where $P_{id}^{(e)} \propto n_e E_F \propto n_e^\gamma$ is the
degenerate electron pressure. 
In analogy with the thermal correction to the electric field, the
correction to the electric field due to Coulomb physics scales as $n_i
E_{\rm C}/n_e E_F$, where $E_{\rm C}$ is the Coulomb energy of the
ions.

Equation (\ref{eq:full E-field}) is the central result of this section
and we now briefly discuss its essential physics.  When $A_1/Z_1 \ne
A_2/Z_2$, i.e., dissimilar $\mu_e$, the ion scale height is determined
by $k_BT/m_p g$ which is much smaller than the electron (pressure)
scale height $l_e = E_F/m_p g$ in the strongly degenerate part of the
atmosphere.  When $A_1/Z_1 \approx A_2/Z_2$, i.e., similar $\mu_e$, then the
zeroth order field produces no differentiating force (modulo thermal/mass defect corrections).  Moreover, the
thermal correction to the electric field produces an ion scale height
$\approx l_e$ as shown earlier.  Mass defections corrections produce ion scale heights that are similarly large.  While this results in a decreasing
fraction for the lighter ion going deeper into the envelope, the
column of this lighter ion increases as discussed earlier (and increase really fast if we include mass defect corrections).  Including
Coulomb physics causes the ion scale height to become $l_e/\Gamma_i$,
much smaller than $l_e$ for large $\Gamma$, but is still larger than
the ion scale height for dissimilar $A/Z$ by $E_{\rm C}/E_F$.  Both
concentration and number density for the lighter ion decrease with
increasing depth.

We now utilize the discussion above to estimate $y_{\rm
  cut}$. Plugging equation (\ref{eq:simple E-field}) into hydrostatic
balance (\ref{eq:hb ions}) for a trace, 2, and a background, 1, we
find for relativistic degenerate electrons ($\gamma = 4/3$) that
\begin{eqnarray}
\label{eq:background}
\frac {\partial\ln n_1} {\partial r} &=& \frac{m_p g}{E_F} \Gamma_e Z_1^{5/3}, \\
\label{eq:trace}
\frac {\partial\ln n_2} {\partial r} &=& \frac{m_p g}{E_F} \Gamma_e Z_2^{5/3}.
\end{eqnarray}
Subtracting (\ref{eq:background}) from (\ref{eq:trace}) and noting
that $\partial\ln P/\partial r = \partial\ln y/\partial r = -\rho g/P = - 4Am_pg/ZE_F$,\footnote{The
  factor of 4 comes from noting that $P\approx P_e$, i.e.,
  relativistic degenerate electrons dominate the pressure, and $P_e =
  n_e E_F/4$.} we find
\begin{equation}\label{eq:coulomb power law}
 \frac {\partial \ln n_2/n_1} {\partial \ln y} = \frac 1 4 \frac{Z}{A} \Gamma_e\left(
 Z_2^{5/3} - Z_1^{5/3} \right)
\end{equation}
Plugging in values for a He/C envelope appropriate for Figure
\ref{fig:HeCstructure} with base temperature $T_b = 1\times 10^8$ K (see \S\ref{sec:He DNB} for a discussion of the base temperature), we find that
${d\ln (n_2/n_1)}/{d\ln y} = -1$ at $\rho = \rho_{\rm cut} \approx 5
\times 10^6\,{\rm g\,cm}^{-3}$ for relativistic degenerate electrons,
which gives $y_{\rm cut} \approx 2 \times 10^9\,{\rm g\,cm}^{-2}$.

The results of this section show that stratification occurs even for
elements with the same $\mu_e$.  This has important implications in
high gravity environments like white dwarfs and NSs beyond the problem
of He DNB.

\section{Diffusive Nuclear Burning of Helium}\label{sec:He DNB}

Combining the results of \S \ref{sec:efield}, the thermal profile of
the envelope, rates of He burning, and rates of diffusion, we now
calculate the rate of nuclear burning of He in NS envelopes.  For the
thermal profile, we solve the heat diffusion equation for a constant
flux envelope in radiative equilibrium:
\begin{equation}
\frac {\partial T} {\partial r} = -\frac {3 \kappa \rho} {16 T^3}T_e^4,
\label{eq:flux}
\end{equation}
where $\kappa$ is the opacity. Starting from the photosphere (with optical depth, $\tau = 2/3$), we integrate the plane-parallel envelope inward assuming constant flux to a depth of $y=y_b \equiv 10^{14}\,{\rm g\,cm}^{-2}$, which we call the base of the envelope. 
As in paper II, we use the tabulated conductivities of Potekhin et al. (1999) and the radiative opacities and the equation of state of Potekhin \& Yakovlev (2001), which is applicable to all magnetic field strengths (see paper II for more details). 

Potekhin et al. (1997) performed a similar calculation down to a density of $\rho=\rho_b\equiv 10^{10}\,{\rm g\,cm}^{-3}$, which they referred to as the base of the envelope.  These two definitions are roughly equivalent to each other, namely that $\rho(y_b) \approx 10^{10}\,{\rm g\,cm}^{-3}$.  We define the temperature at $y_b$ as the base temperature, $T_b$.  For a NS in thermal equilibrium, the base temperature and the core temperature, $T_c$ are equal to each other as the temperature profile is nearly isothermal due to the large conductivities.  For the most part, $T_b$ and $T_c$ can be use interchangable, except for the very early cooling history of the NS (age $\lesssim 100$ yrs), when the NS is not in thermal equilibrium and $T_b\neq T_c$. This point will be important in our discussion of Figure \ref{fig:He burning lifetime}.

For the diffusion coefficient of the trace He, we use the results of
paper II (also see Brown, Bildsten \& Chang 2002), where we found:
\begin{equation}\label{eq:diff_coeff}
  \D \approx 10^{-3} \frac{A_1^{0.1} T_6^{1.3}}{Z_1^{1.3}
    Z_2^{0.3} \rho_5^{0.6}} \,{\rm cm}^2\,\sec^{-1},
\end{equation}
where $T_6 = T/10^6\,{\rm K}$ and $\rho_5 = \rho/10^5 \,{\rm
  g\,cm}^{-3}$.  As discussed in paper II, equation
(\ref{eq:diff_coeff}) is reasonably accurate in the liquid regime of
the background C, $1 < \Gamma < \Gamma_M$, where $\Gamma_M = 175$ defines the melting point of the crystalline phase (Potekhin \& Chabrier 2000).  For $\Gamma > \Gamma_M$, the
material is a crystalline and we assume that there is no diffusion, though we never reach this point in our parameter regime.

%\subsection{He Burning Processes}

We now discuss the nuclear processes that consume He.  For our nuclear
reaction rates, we have utilized both the
NACRE\footnote{http://pntpm.ulb.ac.be/Nacre/nacre.htm}(Angulo et
al. 1999) and REACLIB\footnote{http://nucastro.org/reaclib.html}
compilation of nuclear reaction rates and experimental values (see
nucastro.org).  We have not included electron screening in our
calculations, though the effect of screening would be to increase the
rate of burning.  For triple-$\alpha$ reactions, we use the fit in
Nomoto, Thielemann, \& Miyaji (1985), but have found this rate to be
typically very small for the parameters of interest because of the
reduced number density of He in the burning region (note the reduced
number fraction of He in Figure \ref{fig:HeCstructure}). Hence, we no
longer discuss triple-$\alpha$ reactions.

The local rate of He burning is
\begin{equation}
\dot{y}_{\rm He} \equiv \frac{y_{\rm He}}{\tau_{\rm He, col}} = \int
dz \frac{ 4 n_{\rm He} m_p}{ \tau_{\rm He}(n_{\rm He},n_{\rm \alpha-cap},T)},
\label{eq:rate}
\end{equation}
where $z$ is the depth and $n_{\rm \alpha-cap}$ is the density of a
He-capturing substrate.

\begin{figure}
%  \epsscale{0.9}
  \plotone{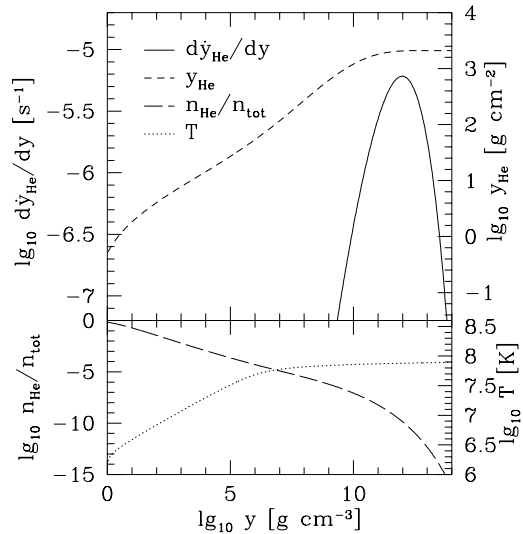}
  \caption{Profile of He burning (solid line) and $y_{\rm He}$
    (short-dashed line) as a function of column with $T_e = 1.25
    \times 10^6\,{\rm K}$ (upper panel).  Shown in the lower panel are
    temperature (dotted line) and He abundance (long-dashed line).
    The base temperature is $7.8\times 10^7$ K, photospheric He
    abundance is 50\%, and the total He column depth is $y_{\rm He}
    \approx 2100\,{\rm g\,cm}^{-2}$.  At this base temperature, the
    rate of He burning is limited by the rate of alpha capture onto
    carbon, i.e., the nuclear limited regime of DNB. The He tail which
    penetrates into the C remains in diffusive equilibrium as the rate
    of alpha capture is slow.}
  \label{fig:he burning profile 1}
\end{figure}

\begin{figure}
%  \epsscale{0.9}
  \plotone{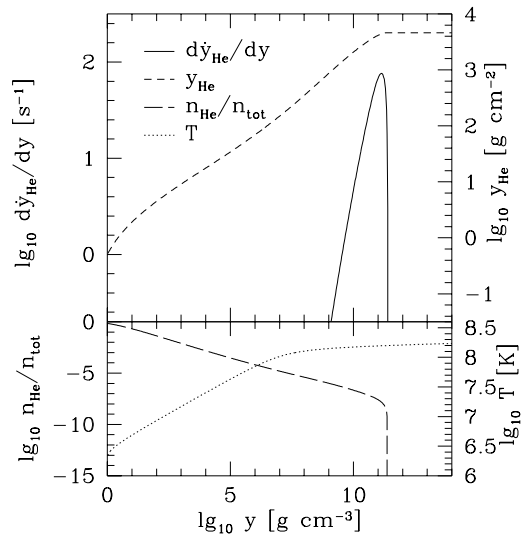}
  \caption{Same as Figure \ref{fig:he burning profile 1} but with $T_e
    = 2 \times 10^6\,{\rm K}$.  The base temperature is $T_b =
    1.7\times 10^8$ K, photospheric He abundance is 50\%, and the
    total He column depth is $y_{\rm He} \approx 4600\,{\rm
      g\,cm}^{-2}$. Unlike Figure \ref{fig:he burning profile 1}, the
    rate of He burning is limited not by the rate of alpha capture
    onto carbon, but on the rate of He diffusion down to the burning
    layer, i.e., the diffusion limited regime of DNB. The He tail
    which penetrates into the C is not in diffusive equilibrium below
    the burning layer.}
  \label{fig:he burning profile 2}
\end{figure}

In the top panels of Figures \ref{fig:he burning profile 1} and
\ref{fig:he burning profile 2}, we plot the rate of He burning (solid
lines) as a function of column, i.e., $d\dot{y}_{\rm He}/d y$ for
effective temperatures of $T_e = 1.25\times 10^6\,{\rm K}$ and
$2\times 10^6\,{\rm K}$ with photospheric He abundances of $50\%$.  We
also plot the $y_{\rm He}$ (short-dashed lines) as a function of
column.  Plotted in the bottom panels of these two figures is the
background temperature profile (dotted lines) and the composition
profile (long-dashed lines) of the envelope.  Figure \ref{fig:he
  burning profile 1} and \ref{fig:he burning profile 2} represent He
DNB in the nuclear limited regime and the diffusion limited regime
respectively (see \S\ref{sec:physics} and paper II for a detailed
discussion of the nuclear limited and diffusion limited regimes of
DNB).  In both cases, the peak of the burning is set by a combination
of a rising temperature profile and a rapidly falling He
concentration.  The cutoff in the nuclear-limited regime
(Fig. \ref{fig:he burning profile 1}) results from the equilibrium
profile of He embedded in a C background.  Hence, the He profile
remains in diffusive equilibrium as it is slowly consumed.

On the other hand, the sharp dropoff in the He profile in the
diffusion limited regime (Fig.\ref{fig:he burning profile 2}) is not a
result of the equilibrium structure, but rather due to nuclear burning
forcing the He profile out of diffusive equilibrium.  As in the case
of diffusion limited H-DNB discussed in paper II, the rate at which H
is consumed is ultimately limited by the rate at which H diffuses down
to the burning layer.  Below this layer H does not exist as the
burning layer captures all downwardly diffusing protons.  The rate of
nuclear burning is not limited by the rate of nuclear reactions, but
rather by the rate of diffusion down to this boundary.  We see this
same behavior for He at sufficiently high temperatures.

To calculate the case of diffusion limited He DNB, we have generalized
the argument of paper II.  The details can be found in paper II, but
for completeness, we write the additional equation to be solved below:
\begin{equation}\label{eq:f equation}
\frac {\partial^2 f_{\rm He}}{\partial z^2} + \frac {\partial f_{\rm He}}{\partial z} \frac {\partial \ln n_{\rm He,0}}{\partial z} = \frac {f_{\rm He}}{\D\tau_{\rm He, nuc}},
\end{equation}
where $n_{\rm He,0}$ is the He number density in the absence of
nuclear reactions, $f_{\rm He}\equiv n_{\rm He}/n_{\rm He,0}$ is the
correction factor in the presence of nuclear reactions, and $n_{\rm
  He}$ is the He number density.  In the regime where equation
(\ref{eq:f equation}) becomes important, the second term on the LHS
becomes small compared to all the other terms.  In other words, the
scale height of $f_{\rm He}$ is small compared to the equilibrium
scale height, $h_{\rm eq} = (\partial\ln n_{\rm He,0} /\partial
z)^{-1}$.  As we found in paper II, equation (\ref{eq:f
  equation}) becomes a diffusion equation in the presence of a nuclear
driven source.

We note that the small difference between the effective temperature
between Figure \ref{fig:he burning profile 1} and \ref{fig:he burning
  profile 2} ($1.25\times 10^6\,{\rm K}$ vs $2\times 10^6\,{\rm K}$)
yields a seven order of magnitude difference in burning rate and He
column lifetime.  This results from the strong temperature sensitivity
of $\alpha$ capture onto C.

We now discuss how the lifetime of a He layer, $\tau_{\rm He} = y_{\rm
  He}/\dot{y}_{\rm He}$, scales with its column, $y_{\rm He}$.  In
Table \ref{table:tau}, we show the photospheric He abundance ($n_{\rm
  He}/n_{\rm tot}$ at the photosphere), He column $y_{\rm He}$ (for He
sitting on top of C), and the He lifetime, $\tau_{\rm He}$, for a NS
with $T_e= 2 \times 10^6$ K.  We note two remarkable things about this
table.  First, the He lifetime is independent of the total column.
This was not the case for H DNB (papers I,II,III). Second, even though
the photospheric abundance of He can be low, the total column of He is
high.  Namely, for a photospheric He abundance of 10\%, the total He
column is $\approx 2500\,{\rm g\, cm}^{-2}$ much greater than the
photospheric column of $y_{\rm ph}\approx 1\,{\rm g\,cm}^{-2}$.

\begin{deluxetable}{c c c}
  \tablecolumns{3}
  \tablewidth{0pt}
  \tablecaption{Photospheric He abundance, He column, and corresponding lifetime for a He/C envelope on a NS with $T_e= 2 \times 10^6$ K.
    \label{table:tau}}

  \tablehead{\colhead{He photospheric abundance} & \colhead{$\lg_{10} y_{\rm He}$ [${\rm g\,cm}^{-2}$]} & \colhead{$\lg_{10} \tau_{\rm He}$
    [yrs]}}        
        \startdata
        0.1 & 2.888 & 1.947 \\
        0.3 & 3.397 & 1.947 \\
        0.5 & 3.664 & 1.947 \\
        0.9 & 4.144 & 1.947 \\
        0.999 & 4.851  & 1.947 \\
        0.999999 & 5.848 & 1.949
        \enddata
\end{deluxetable}

We now demonstrate that the independence of $\tau_{\rm He}$ and
$y_{\rm He}$ and $y_{\rm He} \gg y_{\rm ph}$ are due to the same
physics.  The reason why $y_{\rm He} \gg y_{\rm ph}$ is because the
bulk of the He is not near the photosphere, but rather deep in the
envelope.  We show this in the top panels of Figures \ref{fig:he
  burning profile 1} and \ref{fig:he burning profile 2} where we plot
$y_{\rm He}$ (short-dashed lines) as a function of $y$.  Note that the
bulk of the He does not reside near the surface, but increases
substantially until is asymptotes at a column of $y\approx
y_{\rm cut}$.  

We now explain why the bulk of He resides near $y_{\rm cut}$. In paper
I, we showed that the total lifetime of H on a proton capturing
substrate is dominated by the time it takes to remove a photospheric H
column. Namely, we found that the H lifetime, $\tau_{\rm H}$, scales
with the H column as $\tau_{\rm H} \propto y_{\rm H}^{1+\delta}$,
where $\delta = A_2(Z_1 + 1)/A_1 - Z_2 - 1$, where $1$ and $2$ denote
the background and trace ion species respectively, i.e.,
$1+\delta=-5/12 < 0$ for a H on C envelope (paper I).  As we discussed
in \S 5.2 of paper I, this scaling arises from the power-law falloff
of H abundance in a nondegenerate atmosphere, which is $n_{\rm H}/n_{\rm
  tot} \propto y^{\delta} = y^{-17/12}$ (see eq. [31] of paper I).
For He on C, the He abundance follows $n_{\rm He}/n_{\rm tot} \propto
y^{\delta} = y^{-2/3}$.  The H abundance on a C substrate falls off
steeper than $y^{-1}$, whereas the He abundance on a C substrate falls
off shallower than $y^{-1}$.  Thus, the H column is capped at the H/C
boundary, the column where the number density of H and C are equal,
but the He column continues to increase substantially below the He/C
boundary.  Indeed, one can show that the He column always increases
below the He/substrate boundary for any substrate with $A/Z \approx 2$.
This increasing He column carries through to the Coulomb case as
discussed earlier in \S\ref{sec:coulomb}.  Indeed, the He column
continues to rise until it is capped at a column of $y_{\rm cut}$ as
we argued in \S\ref{sec:thermal}.

We now note that the burning layer is close to the region where He
begins to get cut off, $y_{\rm burn} \sim y_{\rm cut}$ (see for
instance, Figures \ref{fig:he burning profile 1} and \ref{fig:he
  burning profile 2}).  If we say that they are the same, we are left
with the result that the He abundance in the burning layer and the
total He column are {\it linearly} related to each over.  Since the
rate of He DNB (via captures onto C) is linearly related to the He
abundance (and hence $\dot{y}_{\rm He} \propto y_{\rm He}$), we find
that the lifetime of the He layer, $\tau_{\rm He} = y_{\rm
  He}/\dot{y}_{\rm He}$ is independent of $y_{\rm He}$ as demonstrated
in Table \ref{table:tau}, explaining the independence of $\tau_{\rm
  He}$ and $y_{\rm He}$.

\begin{figure*}
%  \epsscale{0.9}
  \plottwo{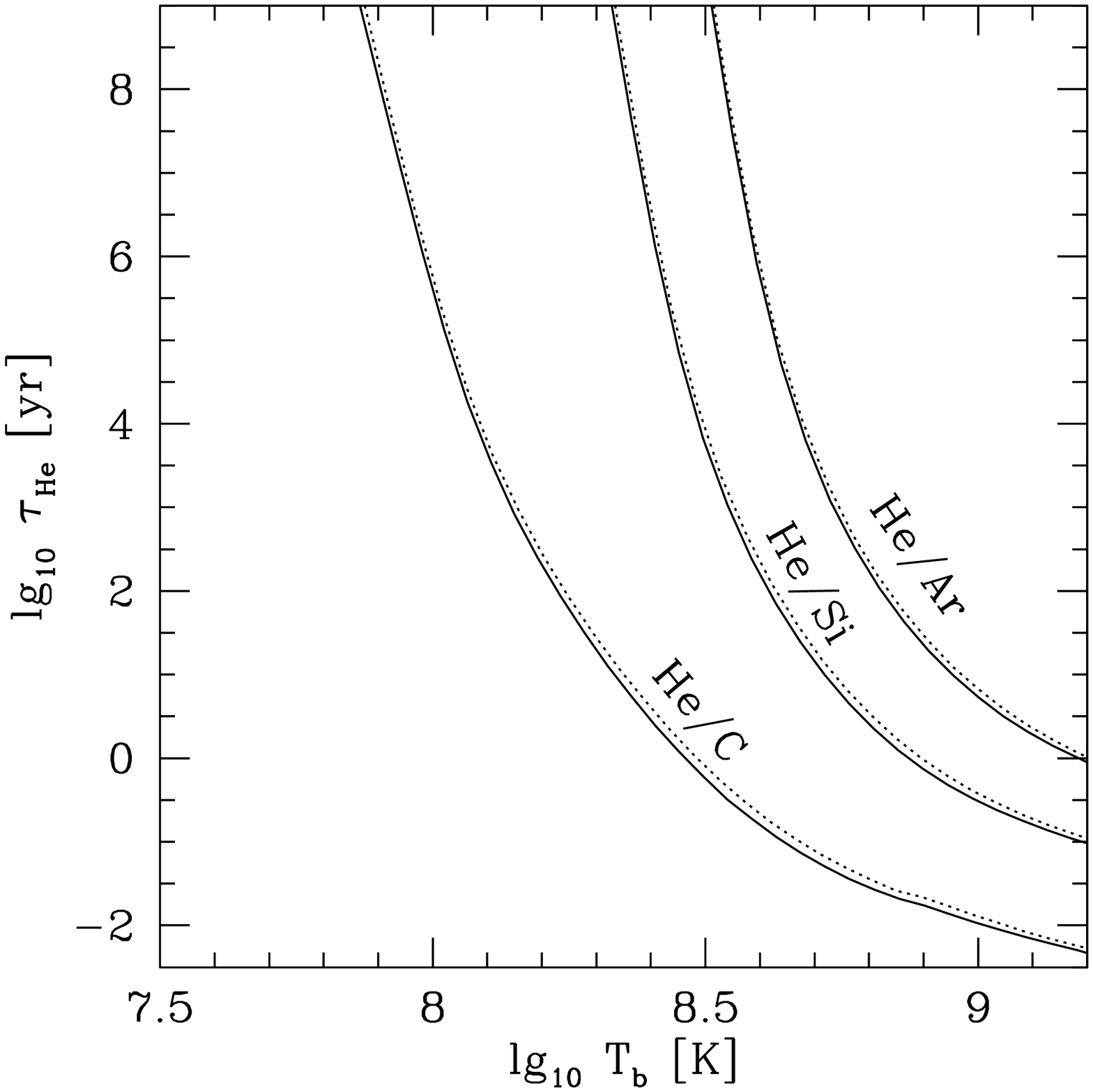}{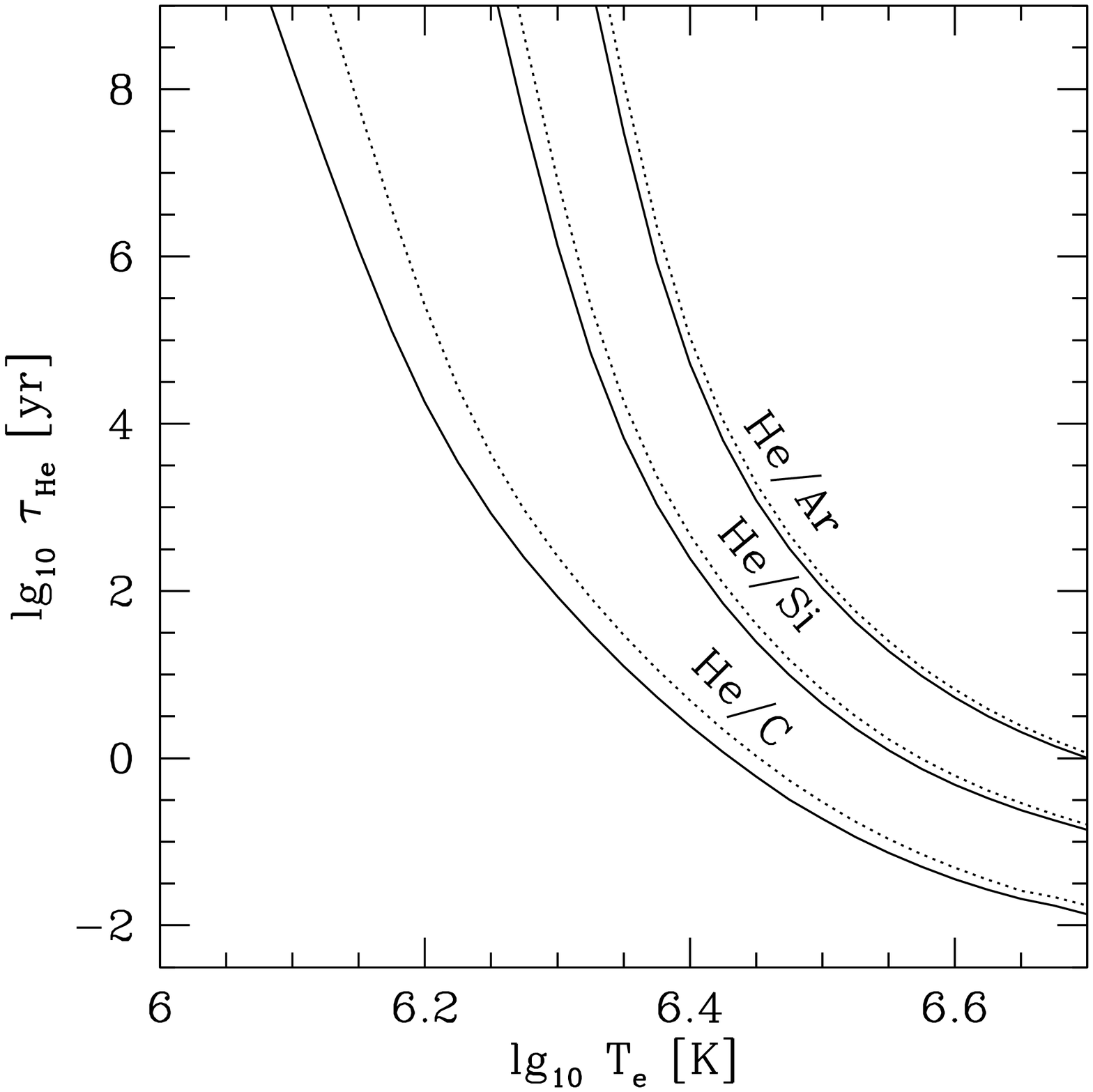}
  \caption{Lifetime for He sitting on top of C, Si, or Ar envelope as
    a function of base temperature, $T_b$ (left plot), and as a
    function of effective temperature, $T_e$ (right plot) for $B=0$ (solid lines) and $B=10^{12}$ G (dotted lines).  For
    sufficiently high $T_b$ ($T_e$), i.e., $T_b \gtrsim 1.6 \times
    10^8$ K ($T_e \gtrsim 2\times 10^6$ K), the lifetime of He against
    capture on C is very short ($ \lesssim 10^2$ yrs).  As these
    $T_b$'s are reached during the early cooling history of the NS,
    primordial He should be burned off. At low temperatures, there is
    a steep power-law dependence, as would be expected for DNB in the
    nuclear-limited regime.  At higher temperature, the dependence on
    $T_b$ becomes shallower, indicative of DNB in the diffusion
    limited regime. }
  \label{fig:He burning lifetime col}
\end{figure*}

\begin{figure}
%  \epsscale{0.9}
  \plotone{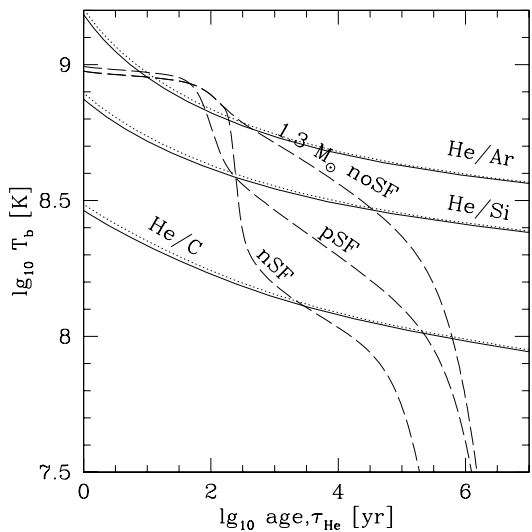}
  \caption{Base temperature as a function of He lifetime/age.
    Overlayed are cooling tracks (long-dashed lines) for various NS
    models (Potekhin et al. 2003) including standard cooling (no
    superfluidity, modified Urca cooling represented by $1.3\Msun$
    noSF), proton superfluidity (pSF) and triplet-state core neutron
    superfluidity (nSF).  Note that for sufficiently young NS or
    sufficiently slow cooling, $\tau_{\rm He} < $ age, which indicates
    that any He on the NS will be consumed by DNB.}
  \label{fig:He burning lifetime}
\end{figure}

In Figure \ref{fig:He burning lifetime col}, we show the He lifetime,
$\tau_{\rm He}$, for a range of $\alpha$-capturing material as a
function of base temperature, $T_b$ (left), effective temperature,
$T_e$, and $B=0$ (solid lines).  We also plot (dotted lines) the same case but for a radial
B-field of $B=10^{12}$ G.  Magnetic fields up to pulsar strengths do
not make a significant difference.  For sufficently large base
temperatures, the lifetime to deplete the atmosphere of He is a few
weeks.

To compare the lifetime of a He layer with the typical cooling history
of a NS, we plot in Figure \ref{fig:He burning lifetime} a few
representative cooling tracks (long-dashed lines) and overlay the
$\tau_{\rm He}$-$T_b$ relation shown in Figure \ref{fig:He burning
  lifetime col}.  Here, we have plotted $T_b$, which during the very early history (age $\lesssim 100$ yrs) of the NS $\neq T_c$.  Only after the cooling wave sweeps through the star, which can be seen by the sudden drop in $T_b$ at $\sim 100$ yrs, is the NS in thermal equilibrium.  As we have done in paper II, we include a standard
cooling track which presumes modified URCA cooling on a 1.3 \Msun\ NS,
another with core proton superfluidity (Potekhin et al. 2003), and one
with triplet-state neutron superfluidity in the core with a maximum critical
temperature of $8\times 10^8 \ {\rm K}$ (calculated by D. G. Yakovlev
previously for paper II). 

The remarkable aspect of this plot is that independent of the cooling
model assumed (where we have taken a few representative models that
span the range from standard cooling to fast cooling), the lifetime of
a He atmosphere is short compared to the cooling age (during its early
history). For instance, He on a C envelope will be depleted even in
the fast cooling case, if it was deposited up to 1000 yrs after
formation.  For slowly cooling models, this can extend up to almost 1
Myr.  For a much heavier material, i.e., Ar, the lifetime is still
short $\sim 100$ yrs for both fast and slow cooling.  Thus, primordial
He should be depleted on NS atmospheres during it early cooling
history.

As we close this section, we note that the lack of observed light
elements on NS photosphere may not indicate that there are no light
elements present.  As we discussed above, in diffusive equilibrium,
the bulk of the He resides near $y_{\rm cut}$.  What this implies is
that while the photospheric abundance of He could be low, the total He
column could be large (see Table \ref{table:tau}).  This result has
implications for the subsequent accretion or production of light
elements on NS surfaces.  For instance, suppose a sudden accretion or
spallation event produces a total He column of $y_{\rm He} \approx
10^3\,{\rm g\,cm}^{-2}$ on the NS surface, which previously consisted
of C.  The NS surface would initially appear to be composed of light
elements.  Over time, as the He diffuses downward to $y_{\rm cut}$, the
photospheric abundance of He drops and the NS surface appears more
C-like.

The timescale for this mixing is the diffusion time down to $y_{\rm
  cut}$ for He on C:
\begin{equation}
\tau_{\rm diff,cut} \approx \frac {h_{\rm p,cut}^2}{\D} \approx 2
\,\rho_{7}^{1.3}T_8^{1.3}\,{\rm yrs},
\end{equation}
where $h_{\rm p,cut}$ is the pressure scale height at $y_{\rm cut}$,
$\rho_7 = \rho/10^7\,{\rm g\,cm}^{-3}$, and $T_8 = T/10^8$ K.  This
timescale is enticingly similar to the timescale of the observed
spectral variations in RX J0720.4-3125 (de Vries et al. 2004; Vink et
al. 2004; Haberl et al. 2006; Hohle et al. 2009).  Haberl et
al. (2006) proposed that NS precession may be responsible for RX
J0720.4-3125's observed spectral hardening and subsequent softening.
However, we speculate that slow evolution of the NS surface over a
period of years may be the culprit. Hydrogen on RX J0720.4-3125 is
consumed by DNB, whereas He would diffusively mix down to $y_{\rm
  cut}$.  A more detailed study of this process of diffusive mixing
for He is needed to make a more quantitative comparions, but such a
calculation is well beyond the scope of this paper.

%\subsection{Magnetic Field Effects on the Rate of He DNB}\label{sec:magnetic}

\section{Conclusions}\label{sec:implications}

We have now extended our prior work on H DNB to He. We now include thermal, mass defect, and Coulomb corrections to the electric field to
calculate the structure of He on C (or any other $A/Z\approx 2$ material) in
diffusive equilibrium.  We show that the bulk of He in diffusive
equilibrium sits at a layer $y_{\rm cut}$ and show that as result of
this the lifetime of a He on a NS is independent of its column.  We
demonstrate that young NSs undergo a phase where He is consumed via
DNB (see especially Figure \ref{fig:He burning lifetime}). Combined
with the results of paper I, II, and III (see Figure 8 of paper II),
we conclude that all primordial H and He are depleted on the surfaces
of NS in the absence of external sources.  In addition, the effects of
magnetic field, as discussed in \S\ref{sec:He DNB} (also see papers II
and III for H DNB), do little to change this basic result. We require
the presence of H and He capturing material in NS envelopes, which the
initial nuclear evolution of a cooling NS produces (Hoffman \& Heyl
2009), i.e., proton and He capturing elements as light as Si.
Finally, we comment on the timescale for He on C to reach diffusive
equilibrium and suggest that such an evolution may explain the
observed long-term spectral variation seen in RX J0720.4-3125.

Since H and He are consumed during the NS's early history, we expect
that the NS surfaces will be dominated by mid-Z elements such as C, N,
and O.  The evidence for mid-Z elemental compositions on NSs is
sparse. Observations of 1E1207.4-5209 by Sanwal et al (2002) and
Mereghetti et al. (2002) found X-ray spectral lines at 0.7 keV and 1.4
keV with additional spectral features at 2.1 and 2.8 keV, which have
been disputed (Mori et al. 2005).  Subsequent modeling by Hailey \&
Mori (2002), Mori \& Hailey (2006), and Mori \& Ho (2007) suggested that its atmosphere
consists of mid-Z elements like O or Ne with a magnetic field strength
of $6\times 10^{11}\,{\rm G}$.  Given the age of 1E1207-52 of $\approx
7$ kyrs (Roger et al. 1988), we found that H is easily consumed (paper
II).  Using the results of this paper, we suggest that He could have
been consumed during the early cooling history of the NS exposing the
underlying O or Ne.  Subsequent accretion was O-rich and not subject
to spallation or that the strong pulsar wind prevents accretion.  We
also note that removal via pulsar winds remains a strong possibility.

However, more recent work by Gotthelf \& Halpern (2007) challenges the
intepretation that these spectral lines on 1E1207-52 are due to mid-Z
materials.  Rather the spindown rate which Gotthelf \& Halpern (2007)
found suggest a B-field strength ($B < 3\times 10^{11}$ G), consistent
with electron cyclotron lines for the higher energy (1.4 keV) line.

Currently the best evidence for mid-Z atmospheres comes from Ho \&
Heinke (2009).  They recently argued that the surface of the NS in the
center of Cassiopeia A consist of a C atmosphere.  Their modeling of
archival observations of this compact X-ray source showed that a
$T_{\rm e} = 1.8\times 10^6$ K carbon atmosphere NS with a low
magnetic field provides both a good fit to the observed spectrum and
is consistent with the theoretical expectation of the radii of NSs,
i.e., $R = 12-14$ km. Additional observations of other young NS may
provide further evidence of mid-Z elements in their atmospheres.

We should mention that even though DNB should deplete the atmospheres
of young NSs of any primordial H or He, subsequent accretion with or
without spallation (Bildsten, Salpeter, \& Wasserman 1992) could still
lead to a H/He atmosphere in spite of DNB.  As the mass of the
photosphere is low and the timescales still fairly long, even a small
amount of accretion will pollute the photosphere.  A quick estimate of
the amount of accretion needed to overwhelm DNB is
\begin{equation}\label{eq:mdot}
\dot{M} > 10^{-30} \left(\frac {y_{\rm ph}}{1 \,{\rm g\,cm}^{-2}}\right) \left(\frac {\tau_{\rm
  ph}}{100\,{\rm yrs}}\right) \,M_{\odot}\,{\rm yr}^{-1}.  
\end{equation}
A pulsar wind from the NS may suppress this accretion, but the amount
of suppression has to be fairly strong as indicated by equation
(\ref{eq:mdot}).  
 
Armed with the results of this work and prior efforts, we expect that
primordial H and He are depleted, and so any observed H or He on the
surfaces of these NS must be due to subsequent accretion (with or
without spallation).  If this subsequent accretion can be stopped or
prevented via a pulsar wind, for instance, the underlying mid-Z
material would be exposed.

\acknowledgements

We thank W. Ho and M. van Kerkwijk for useful discussions, B. Paxton for pointing out nuastro.org and general awesomeness, and the anonymous referee for a thorough reading of the manuscript and making excellent comments.
P.C. is supported by the Canadian Institute for Theoretical
Astrophysics.  P.A. is an Alfred P. Sloan Fellow, and acknowledges
support from the Fund for Excellence in Science and Technology from
the University of Virginia. P.A. also acknowledges support from NASA
grant NNX09AF98G and NSF grant AST-0908873.  This work was supported by
the National Science Foundation under grants PHY 05-51164 and AST
07-07633. This research has made use of NASA's Astrophysics Data System.\\

%\begin{references}
\references 

\noindent
Angulo, C., et al. 1999, Nuclear Physics A, 656, 3 

\noindent
Bildsten, L., Salpeter, E.~E. \& Wasserman, I. 1992, \apj, 384, 143

\noindent
Blaes, O.~M., Blandford, R.~D., Madau, P., \& Yan, L.\ 1992, \apj, 399, 634

\noindent
Brown, E.F., Bildsten, L. \& Chang, P. 2002, \apj, 574, 920

\noindent
Burwitz, V., Zavlin, V.~E., Neuh{\"a}user, R., Predehl, P., Tr{\"u}mper, J., \& Brinkman, A.~C.\ 2001, \aap, 379, L35 

\noindent
Burwitz, V., Haberl, F., Neuh{\"a}user, R., Predehl, P., Tr{\"u}mper, J., \& Zavlin, V.~E.\ 2003, \aap, 399, 1109 

\noindent
Chabrier, G. \& Potekhin, A.~Y. 1998, \pre, 51, 4941

\noindent
Chang, P. \& Bildsten, L. 2003, \apj, 585, 464 (paper I)

\noindent
Chang, P. \& Bildsten, L. 2004, \apj, 605, 830 (paper II)

\noindent
Chang, P., Arras, P. \& Bildsten, L. 2004, \apj, 616, 147 (paper III)

\noindent
Chiu, H.~Y., Salpeter, E.~E. 1964, \prl, 12, 413

%\noindent
%{Clayton}, D. 1983, \textit{Introduction to Stellar Structure and Nucleosynthesis} (Chicago: University of Chicago Press)

\noindent
De Blasio, F.~V. 2000, \aap, 353, 1129

\noindent
de Vries, C.~P., Vink, J., M{\'e}ndez, M., \& Verbunt, F.\ 2004, \aap, 415, L31 

\noindent
Gotthelf, E.~V., \& Halpern, J.~P.\ 2007, \apjl, 664, L35 

\noindent
Hailey, C.~J. \& Mori, K. 2002, \apj, 578, L133

\noindent
Haberl, F., Turolla, R., de Vries, C.~P., Zane, S., Vink, J., M{\'e}ndez, M., \& Verbunt, F.\ 2006, \aap, 451, L17 

\noindent
Hameury, J.~M., Heyvaerts, J., \& Bonazzola, S.\ 1983, \aap, 121, 259 

\noindent
Ho, W. G. G. \& Heinke, C. O. 2009, \nat, 462, 71

\noindent
Ho, W.~C.~G., Kaplan, D.~L., 
Chang, P., van Adelsberg, M., \& Potekhin, A.~Y.\ 2007, \mnras, 375, 821 

\noindent
Hoffman, K., \& Heyl, J. 2009, \mnras, 400, 1986 

\noindent
Hohle, M.~M., Haberl, F., Vink, J., Turolla, R., Hambaryan, V., Zane, S., de Vries, C.~P., \& M{\'e}ndez, M.\ 2009, \aap, 498, 811 

\noindent
Landau, L. D. \& Lifshitz, E. M. 1980, \textit{Statistical Physics: 3rd Edition Part 1} (Oxford: Butterworth-Heinemann)

\noindent
Mereghetti, S., De Luca, A., Caraveo, P.~A., Becker, W., Mignani, R. \&
Bignami, G.~F. 2002, \apj, 581, 1280

\noindent
Mori, K., \& Hailey, C.~J.\ 2006, \apj, 648, 1139 

\noindent
Mori, K., Chonko, J.~C., \& Hailey, C.~J.\ 2005, \apj, 631, 1082 

\noindent
Mori, K., \& Ho, W. C. G. 2007, \mnras, 377, 905

\noindent
Nomoto, K., Thielemann, F.-K., \& Miyaji, S.\ 1985, \aap, 149, 239 
%\noindent
%Pavlov, G. G., Zavlin, V. E., \& Sanwal, D. 2002, in Proc. 270th Heraeus 
%Seminar on Neutron Stars, Pulsars and Supernova Remnants, ed. W. Becker, 
%H. Lesch, \& J. Trümper (MPE Rep. 278; Garching: MPE)
\noindent
Pons, J.~A., Walter, 
F.~M., Lattimer, J.~M., Prakash, M., Neuh{\"a}user, R., 
\& An, P.\ 2002, \apj, 564, 981 

\noindent
{Potekhin}, A.~Y., {Baiko}, D.~A., {Haensel}, P., \& {Yakovlev}, D.~G. 1999,
  \aap, 346, 345

\noindent
Potekhin, A.~Y. \& Chabrier, G.  2000, \pre, 62, 8554

\noindent
Potekhin, A.~Y., Chabrier, G., \& Rogers, F.~J.\ 2009, \pre, 79, 016411

%\noindent
%Potekhin, A.~Y., Chabrier, G. \& Shibanov, Y.~A.  1999, \pre, 60, 2193

\noindent
{Potekhin}, A.~Y., {Yakovlev}, D.~G., Chabrier, G., Gnedin,
O.~Y. 2003, \apj, 594, 404

\noindent
{Potekhin}, A.~Y. \& {Yakovlev}, D.~G. 2001, \aap, 374, 213

\noindent 
Roger, R.~S., Milne, D.~K., Kesteven, M.~J., Wellington, K.~J., \&
Haynes, R.~F.\ 1988, \apj, 332, 940

\noindent
Rosen, L.~C., 1968, \apss, 1, 372

\noindent
Sanwal, D., Pavlov, G.~G., Zavlin, V.~E. \& Teter, M.~A. 2002, \apj, 571, L61

%\noindent
%Ventura, J. \& Potekhin, A.~Y. 2002, in The Neutron Star--Black Hole
%Connection, ed. C.~Kouveliotou, J.~Ventura, \& E.~{van den Heuvel}, Vol. 567,
%NATO ASI sec.~C (Dordrecht: Kluwer), 393, astro-ph/0104003

\noindent
Vink, J., de Vries, C.~P., M{\'e}ndez, M., \& Verbunt, F.\ 2004, \apjl, 609, L75 

\noindent
Yakovlev, D.G., Kaminker, A.D., \& Gnedin, O.Y. 2001, \aap, 379, 5

%\end{references}

\end{document}